\documentclass[11pt,usletter]{article}

\usepackage[utf8]{inputenc}
\usepackage{jheppub,amsmath,amssymb,mathabx}
\usepackage{bm,bbm}

\allowdisplaybreaks[3]

\newcommand{\prg}{\mathrm{pr}}
\newcommand{\mpl}{M_\mathrm{Pl}}
\newcommand{\op}{\Omega_\mathrm{p}}
\newcommand{\rv}{r_\mathrm{v}}
\newcommand{\rbox}{r_\mathrm{box}}
\newcommand{\diff}[1]{\mathrm{d}{#1}\,}

\newcommand{\p}{\partial}

\newcommand{\nn}{\nonumber}

\newcommand{\mupn}{^\mu_{\, \nu}}

\newcommand{\ie}{i.e.} 
\newcommand{\eg}{e.g.}

\newcommand{\Sect}[1]{Section~#1}
\newcommand{\sect}[1]{section~#1}

\newcommand{\app}[1]{appendix~#1}

\newcommand{\eqn}[1]{eq.~#1}

\newcommand{\fig}[1]{figure~#1}
\newcommand{\figs}[1]{figures~#1}

\newcommand{\tbl}[1]{table~#1}

\begin{document}

\title{Scalar Gravitational Radiation from Binaries: \\
Vainshtein Mechanism in Time-dependent Systems}

\author[a]{Furqan Dar}
\affiliation[a]{Department of Physics, Kenyon College, 201 N College Rd, Gambier, OH 43022, USA}
\author[b,c]{Claudia de Rham}
\affiliation[b]{Theoretical Physics, Blackett Laboratory, Imperial College, London, SW7 2AZ, UK}
\affiliation[c]{CERCA, Department of Physics, Case Western Reserve University, 10900 Euclid Ave, Cleveland, OH 44106, USA}
\author[c]{J.~Tate Deskins}
\author[a,c]{John T. Giblin Jr.}
\author[b,c]{Andrew J. Tolley}

\emailAdd{c.de-rham@imperial.ac.uk}
\emailAdd{jennings.deskins@case.edu}
\emailAdd{giblinj@kenyon.edu}
\emailAdd{a.tolley@imperial.ac.uk}

\abstract{
We develop a full four-dimensional numerical code to study scalar gravitational radiation emitted from binary systems and probe the Vainshtein mechanism in situations that break the static and spherical symmetry, relevant for binary pulsars as well as black holes and neutron stars binaries. The present study focuses on the cubic Galileon which arises as the decoupling limit of massive theories of gravity.
Limitations associated with the numerical methods prevent us from reaching a physically realistic hierarchy of scales; nevertheless, within this context we observe the same power law scaling of the radiated power as previous analytic estimates, and confirm a strong suppression of the power emitted in the monopole and dipole as compared with quadrupole radiation.
Following the trend to more physically realistic parameters, we confirm the suppression of the power emitted in scalar gravitational radiation and the recovery of  General Relativity with good accuracy.
This paves the way for future numerical work, probing more generic, physically relevant situations and sets of interactions that may exhibit the Vainshtein mechanism.
}


\maketitle

\section{Introduction}
\label{sec:intro}
Understanding the physical origin of the observed accelerated expansion of the Universe has led to an explosion of theoretical dark energy and modified gravity models, which incorporate different types of screening mechanisms \cite{Khoury:2010xi,deRham:2012az,Jain:2013wgs,Brax:2013ida}.
These screening mechanisms provide a means by which fields that are active at cosmological scales, potentially significantly modifying the behavior of the gravitational force, are hidden from solar system/astrophysical/lab tests of gravity. 
A large class of theoretical models rely on the Vainshtein screening mechanism \cite{Vainshtein:1972sx} (see \cite{Babichev:2013usa} for a review) which was originally proposed in the context of massive theories of gravity. The essential features of this screening mechanism are captured in the simpler context of Galileon theories \cite{Nicolis:2008in} and pioneering works on how the Vainshtein mechanism manifests itself in models of massive gravity were presented in \cite{Deffayet:2001uk,Dvali:2002vf,Lue:2003ky,Babichev:2009us,Babichev:2009jt,Babichev:2010jd}.

Galileons, a class of scalar field theories that exhibit the nonlinearly realized Galileon symmetry $\pi \to \pi +v_\mu x^\mu +c$, arise naturally from many theories of dark energy like DGP \cite{Luty:2003vm,Nicolis:2004qq}, Ghost-free massive gravity \cite{deRham:2009rm,deRham:2010gu,deRham:2010ik,deRham:2010kj}, Bigravity models \cite{Hassan:2011zd,Fasiello:2013woa} and other theories of massive gravity \cite{Bergshoeff:2009hq,deRham:2011ca} or higher dimensional gravity \cite{deRham:2010eu}.
The Galileon arises as the helicity zero mode of the hard or soft massive graviton, whose nonlinear interactions dominate over the usual helicity two interactions of General Relativity (GR). 
The Galileon effective theory thus describes the region in which the helicity zero mode may be nonlinear, while the helicity two modes are still in the weak field region. 
It is thus sufficient to work with this effective theory in order to describe systems that principally test weak field Einstein gravity, most notably the orbital decay of binary pulsars \cite{Hulse:1974eb,Taylor:1982zz,Taylor:1989sw}.

Further, these scalar field theories are interesting in their own right as intermediate scale effective field theories \cite{Nicolis:2008in} and may admit many infrared (IR) completions (\eg the covariant Galileon \cite{Deffayet:2009wt} is a distinct IR completion from massive gravity \cite{deRham:2010ik,deRham:2010kj}, having the same decoupling limit). 
These theories automatically incorporate the Vainshtein mechanism in a way that is well understood for static sources \cite{Nicolis:2004qq,Nicolis:2008in}.
However the Vainshtein mechanism is altogether less well understood in time-dependent systems. 
Binary pulsar systems are of particular importance since their orbital decay rate provides ones of the most precise tests of GR and its extensions \cite{Hulse:1974eb,Taylor:1982zz,Taylor:1989sw}. 
Binary black hole, neutron star systems, and possibly black hole-neutron star binaries \cite{Seymour:2018bce} are now equally important as sources of directly observed gravitational waves by Advanced Ligo/Virgo \cite{Abbott:2016blz,TheLIGOScientific:2017qsa}.

In the case of binary pulsars, the sources may be treated as approximately non-relativistic since the orbital velocity is typically small. In this non-relativistic limit it was shown in \cite{deRham:2012fw}, that for the cubic Galileon, based on analytic approximations, that the screening of scalar gravitational radiation from binary pulsars is less effective than it is for static sources.
This is due to the introduction of a new length scale associated with the dynamic time scale of the orbit $\op^{-1}$.
In addition to the usual static Vainshtein suppression, there is an enhancement inversely proportional to the velocity of the orbiting system. 
Although \cite{deRham:2012fw} principally considered binary pulsars, similar expectations would hold if the sources were binary black holes or neutron stars --- at least in the region in which curvatures are small --- as the mechanism by which the Vainshtein mechanism works is largely only sensitive to the mass of the source and not on its precise nature.

In the region of parameter space that is relevant for cosmological purposes, this screening is still strong enough to be below current constraints from binary pulsar systems.
It has been suggested that adding the quartic or quintic Galileon terms, which more naturally arise in the context of (hard) massive gravity theories, may weaken the screening enough to be constraining \cite{deRham:2012fg}. 
But the perturbation theory that worked for the cubic Galileon failed for the quartic and quintic Galileon due to the approximations made \cite{deRham:2012fw,deRham:2012fg} (see also \cite{Berezhiani:2013dca} for a discussion of this point). 
Thus, to explore these systems we require the use of better analytic approximations or numerical methods.

In this work we shall use a full three-dimensional, time stepping numerical code that bypasses the need of any analytic approximation or assumption, and is hence usable for any system irrespective of its symmetry (or absence thereof).
The code, a heavily modified version of \emph{Grid and Bubble Evolver} (GABE) \url{http://cosmo.kenyon.edu/gabe.html} \cite{Child:2013ria}, uses a second order finite differencing scheme on a fixed Cartesian grid and integrates in time with an explicit second  order (or optionally fourth order) Runge-Kutta method.
The major modifications include adding a spherical harmonics power computation module, altered boundary and initial conditions, and changes to how the equations of motion are handled in order to deal with the non-linear equations of the Galileons (see \sect\ref{sec:numerics} for a discussion).
Importantly, by using a Cartesian grid we are not making any assumptions about the symmetry of the system and thus solve the full non-linear equations exactly.
This means that the results are an independent way of computing the power radiated by these binary systems from what was done in \cite{deRham:2012fw,deRham:2012fg}.

As a first step we consider the case of the cubic Galileon (decoupled from gravity) coupled to a binary system, whose trace of the stress energy is simulated by a pair of orbiting localized Gaussians on Keplerian orbits. 
The situation naturally applies to binary pulsars, but can be easily modified to describe black hole binaries or neutron stars, in the regime where they are sufficiently far away that the local metric determined by the helicity two modes between the two black holes/neutron stars remains in the weak field limit\footnote{Although black holes are themselves nonlinear solutions, what is relevant is the contribution to the metric in the vicinity of one black hole generated by the other. For sufficient separations this coupling between the two black holes may be well approximated by a linear analysis, at least for the helicity two modes. Clearly when the black holes and neutron stars are close, \ie close to the merger regime, then these approximations will break down. However in this situation the Galileon decoupling limit approximation would not be valid.}. 
In the absence of the Vainshtein mechanism, the predicted scalar gravitational radiation would be of a comparable magnitude to the tensor radiation of General Relativity, something which would be immediately ruled out by constraints on binary pulsars.  
The Galileon interactions capture the nonlinearities which are expected to suppress the scalar radiation relative to the usual tensor contribution.

The case of the cubic Galileon coupled to a binary system has previously been studied analytically in \cite{deRham:2012fw,Chu:2012kz}, and comparison with the analytic results allows us to probe the accuracy of the code.
Alternatively, the success of the numerical code and its agreement with the analytic results allow us to confirm the validity of the analytic estimations performed in \cite{deRham:2012fw,deRham:2012fg}.
Having confirmed the validity of the code, this now opens up the possibility to extend the analysis to other interactions that exhibit the Vainshtein mechanism (\eg~quartic and quintic Galileon interactions and other kinetic types of interactions as in k-mouflage \cite{Babichev:2009ee}) and to other physically relevant situations.

An interesting feature of the analytic results performed in \cite{deRham:2012fw} is the realization that in the time-dependent system, the Vainshtein suppression for the total power radiated in the scalar (dominated by the quadrupole) goes as $(\op \bar r)^{-1}(\op\rv)^{-3/2}$, instead of the expectation of $(\bar r/ \rv)^{3/2}$ from static sources.
This represents an actual enhancement going as $v^{-5/2}$, since for realistic systems like the Hulse-Taylor pulsar the orbital velocity  $v\sim 10^{-3}$ \cite{Taylor:1989sw}.
In the context of the Hulse-Taylor binary system, the overall Vainshtein suppression is still manifest $(\op \bar r)^{-1}(\op\rv)^{-3/2}\ll 1$ and the overall power emitted in the scalar cubic Galileon is negligible compared to that in GR.
Therefore no observable effects would be detected for physical values of the parameters. 
This situation could change in the future as binary pulsars with different orbital frequencies/eccentricities are discovered and longer time measurements are made.
When considering black hole mergers or other astrophysical binaries, we typically expect these systems to be even more relativistic and therefore the Vainshtein suppression to be more pronounced.
As a proof of principle we may consider a very non-relativistic scenario $v=(\op \bar r)\ll 1$, so that even though one may be inside the Vainshtein radius (that is $(\bar r/ \rv)^{3/2} \ll 1$) the hierarchy of scales is such that the power emitted by a `Vainshtein-ly' screened binary system could be higher than what it would have been in the absence of Vainshtein screening. 
This means the hierarchy of scales involved for that to happen are not representative of any realistic physical system that we know, so at the moment this simply appears as a mathematical possibility which has not been realized in nature. 
Nevertheless it raises the question of whether the Vainshtein screening could in some context amplify the power emitted (see \cite{Ogawa:2018srw} for an interesting scenario where anti-screening has been identified numerically).

A very natural concern is whether this enhancement of the power for sufficiently `slow' binary systems while remaining in the Vainshtein region could be an artifact of the assumptions employed in \cite{deRham:2012fw}, where a hierarchy of scales between the Vainshtein radius and the inverse frequency scale was assumed.
Clearly as $\op \to 0$, there is no notion of power emitted and the system reduces to a static one where the Vainshtein suppression goes as $(\bar r/ \rv)^{3/2}$.
In this work we therefore explore the validity of the analytic results \cite{deRham:2012fw} in a region where the hierarchy of scales may not be very strong $(\op ^{-1} \not\ll \rv)$.
Based on the numerical results (that make no a priori assumptions on the scaling), we recover a scaling of the power emitted which is precisely in agreement with the analytic ones from perturbation theory, and does indeed manifest an enhancement of the power emitted while the source remains within the Vainshtein region.
As the hierarchy of scales tends towards a more physically relevant regime, we observe a scaling of the power that remains again in perfect agreement with the analytic results and that suppresses the overall power emitted in the scalar gravitational radiation, providing a good handle on the Vainshtein screening from cubic Galileons in such systems.

In \sect\ref{sec:Context} we briefly review how Galileons emerge from infrared models of gravity, and summarize the Vainshtein mechanism and the expected power emitted in the Cubic Galileon.
We leave the details of the derivation for \app\ref{app:galrad}, where we first review the method for calculating the power from the effective action and then derive the analytic monopole, dipole, and quadrupole power emitted through scalar gravitational radiation in the cubic Galileon. 
We see that the power radiated is indeed dominated by the quadrupole, while the dipole is suppressed by at least 7 orders of magnitude (see \tbl\ref{table1}).

\Sect\ref{sec:numerics} then focuses on the numerical work we performed with the Galileon theories.
We enumerate the difficulties of the numerical approach due to the many conflicting dynamical scales and the non-linearities associated with the Galileons in binary systems.
We continue on to discuss how the power calculated in the simulation differs from that computed from perturbation theory due to the lack of hierarchy between the size of the source, the inverse frequency scale, and the Vainshtein radius.
We perform tests of the numerics with the well understood Klein-Gordon (free field and thus no Vainshtein mechanism) matching within 1\% of the well-known analytic results when the scales are well resolved by the numerics.
We then continue on to work with the cubic Galileons where we expect a Vainshtein mechanism to affect the amount of power emitted (in a multipole dependent manner).
Despite the lack of strong hierarchy between the size of the source, the inverse frequency scale, and the Vainshtein radius, the scaling of the power with angular velocity and Vainshtein radius  is in good agreement with the analytic results summarized in \sect\ref{sec:Context}.
Finally we summarize our results in \sect\ref{sec:conc}.
We refer to \app\ref{app:rescale} for the rescaling used in the program.

\section{Scalar Gravitational Radiation and Vainshtein}
\label{sec:Context}

Scalar gravitational radiation is a generic feature of modified theories of gravity which include additional degrees of freedom that couple to the trace of the stress-energy. 
The scalar may be a new fundamental degree of freedom, such as in the case of Brans-Dicke theories \cite{Brans:1961sx}, $f(R)$ theories \cite{Carroll:2003wy}, covariant Galileon \cite{Deffayet:2009wt} or Horndeski theories \cite{Horndeski:1974wa}, or they may arise as helicity zero modes of fields of spin-1, 2 or higher. 
Notable in the latter are theories of massive gravity, both soft massive graviton theories such as the Dvali-Gabadadze-Porratti model \cite{Dvali:2000rv} or cascading gravity \cite{deRham:2007rw,deRham:2007xp}, and hard massive graviton theories \cite{deRham:2010ik,deRham:2010kj}\footnote{For a discussion on the distinction between hard and soft massive gravity and observational bounds on massive gravity theories, including those from gravitational wave and pulsar observations see \cite{deRham:2016nuf}.}, as well as generalized Proca theories (see \cite{Heisenberg:2017mzp,Heisenberg:2018vsk}  recent reviews).

Massive gravity theories, both soft and hard, automatically incorporate the Vainshtein mechanism.
The scalar degree of freedom that arises in massive gravity behaves as a Galileon and is made transparent by considering the decoupling limit, $\mpl\to \infty$, $m\to 0$, keeping the scale $\Lambda = (m^2 \mpl)^{1/3}$ fixed, where $m$ is the mass of the graviton \cite{deRham:2010ik,deRham:2010kj}.
Physically this Galileon-like degree of freedom is the helicity zero mode of the massive graviton. 
Conventionally denoted by $\pi$, it is built into the spin-2 field in the manner
\begin{equation}
h_{\mu\nu} = h^E_{\mu\nu} + \pi \eta_{\mu\nu}
\end{equation}
where $h^E_{\mu\nu}$ denotes the Einstein frame metric perturbation, for which there are no quadratic kinetic mixings between $h^E_{\mu\nu}$ and $\pi$. 
It is precisely because of this decomposition that the stress energy coupling $h_{\mu\nu}T^{\mu\nu}$ automatically includes a source for the scalar field $\pi T$, which is the principle source of scalar gravitational radiation. 
This source comes hand in hand with the Galileon interactions \cite{Luty:2003vm,deRham:2010ik,deRham:2010kj} which are responsible for the Vainshtein screening \cite{Nicolis:2004qq}.

In the case of static spherically symmetric sources the Vainshtein screening mechanism is well understood, and is known to apply in a finite region $r < \rv$ outside of any matter source. 
In practice it states that inside the Vainshtein radius $r<\rv$ of some heavy mass source, the coupling of the canonically normalized helicity zero mode to the trace of the stress energy of any additional matter is suppressed $r/\rv$ to the power of a positive exponent which depends on the precise theory. 
Hence the fifth force between two small bodies in the presence of a large mass is suppressed by  $(r/\rv)^q$ with $q>0$.
This suppression will occur even if the source is a black hole, since what matters is the background $\pi$ field generated by the heavy mass source which, for static configurations, is determined by a Birkhoff type theorem to depend only on the total mass of the source.

It is far less well understood how well the Vainshtein mechanism works in situations where there is no spherical symmetry or there is time dependence (see for example \cite{Hiramatsu:2012xj,Brito:2014ifa,Winther:2015pta}). 
In the case of the cubic Galileon with a rotating binary source, an approximate analytic treatment was given in \cite{deRham:2012fw} which led to the conclusion that in addition to the static suppression $(r/\rv)^{3/2}$, there is an enhancement in the emitted radiative power given by an inverse power of the rotation velocity.
The essential steps of the approximate analytic treatment are reviewed in \app\ref{app:galrad}. They begin with the cubic Galileon theory
\begin{align}
	\label{eqn:galaction}
	S = \int \diff{^4 x}\left(-\frac34 (\partial\pi)^2-\frac1{4\Lambda^3} (\partial\pi)^2 \Box \pi + \frac{1}{2\mpl}\pi T \right)\,,
\end{align}
for which the Vainshtein radius associated with a source of mass $M$ is given by
\begin{align}
\label{eqn:rv}	
\rv = \frac1{\Lambda}\left(\frac M{16 \mpl}\right)^{1/3}.
\end{align}
In the absence of the Vainshtein mechanism, \ie in the absence of the Galileon interactions, the power radiated in the quadrupole mode for the free field Klein-Gordon equation with the same source is
\begin{equation}
	\label{eqn:kgp2_mpl}
	P_2^\text{KG} =\frac{1}{45} \frac{M^2}{8\pi \mpl^2} ( \op \bar r)^4\op^2,
\end{equation}
leading to an enhancement of 5\% as compared to the power emitted in GR.
By imposing a Keplerian orbit (so the dependence on $\op$ is manifest) this becomes
\begin{equation}
	\label{eqn:kgp2_pr}
	P_2^\text{KG} = \frac{M}{\bar r} \frac{( \op \bar r)^8}{45}.
\end{equation}

Following the approximate treatment of \cite{deRham:2012fw} (\app\ref{app:galrad}), the power radiated in the cubic Galileon is
\begin{equation}
	P_2^\text{cubic}=\frac{M^2}{8 \pi \mpl^2} \frac{45\times 3^{1/4} \pi^{3/2}}{1024\, \Gamma
   \left(\frac{9}{4}\right)^2} \frac{(\op\bar r)^3}{ (\op\rv)^{3/2}}\op^{2}.
\end{equation}
Comparing to the Klein-Gordon result we find
\begin{equation}
	\label{eqn:cu_power_rat}
	\frac{P^\text{cubic}_2}{P^\text{KG}_2} = \frac{25\times 3^{17/4}\pi^{3/2}}{1024\, \Gamma\left(\frac94\right)^2} (\op\bar r)^{-1} (\op\rv)^{-3/2}.
\end{equation}
Importantly the normal Vainshtein suppression of the Galileon force between the two static objects scales as $(\bar r/\rv)^{3/2}$, whereas in this dynamic system it scales as $(\op\bar r)^{-1} (\op\rv)^{-3/2}$. 
This implies a weakening of the Vainshtein mechanism by the orbital velocity $v^{5/2}=(\op\bar r)^{5/2}$ in comparison to the static case.
Lastly we should note that the above analytic calculation is an approximation, which may only have a limited regime of validity. Simple estimates suggest there are order unity corrections to the source that arise from next order in perturbation theory, which appear as $\sim(\Box \phi)^2/\Lambda^3$. 
At leading order these terms are independent of $\rv$ and $\op$ and thus will only modify the overall proportionality constant and not the scaling of the power with $\op$ and $\rv$. 
The central goal of the present paper is to confirm the essential features of these analytic results through a full four dimensional numerical simulation.

\section{Numerical Methods}
\label{sec:numerics}

Numerical work with time dependent Galileon theories is quite young. 
Most work has been done in the quasi-static limit \cite{Barreira:2013eea,Li:2011vk,Li:2013nua,Khoury:2009tk,Schmidt:2009sv,Schmidt:2009sg,Lue:2004rj,Lue:2003ky} with relevance to N-body simulations or cosmological perturbations. 
This limit does not include the radiative modes, which are our concern here.
Because of the non-linear nature of these theories, questions about the stability of numerical work and even whether it can be simulated at all have arisen \cite{Brito:2014ifa}. 
For example, in \cite{Brito:2014ifa} they perform nonlinear simulations for spherically symmetric configurations. 
The potential onset of Cauchy instabilities can be seen by perturbing around background solutions and determining the form of the effective metric for fluctuations. 
For certain backgrounds the effective metric can have vanishing $Z^{tt}$ component (see \eqn\eqref{eqn:pertquadact}), already noted in \cite{Adams:2006sv} (see \cite{Babichev:2007dw} for a related discussion for k-essence).
If such configurations arise dynamically, then constant $t$ hypersurfaces are no longer good Cauchy surfaces. 
This does not preclude the possibility that other surfaces can act as Cauchy surfaces, for instance those defined by a field dependent coordinate transformation. 
Furthermore there remains the question of whether such configurations would ever arise in a fully quantum effective field theory \cite{Burrage:2011cr,Babichev:2007dw}.

In the present work we perform simulations with the full non-linear equations of motion making no symmetry assumption, and we do not see any issue with Cauchy stability for our simulations. 
This confirms that there is at least a finite set of initial data for which the dynamical evolution of the effective theory is well defined for arbitrarily long times, consistent with previous results. 
The full issue of the well-posedness of the Galileon system is beyond the scope of this paper. However, we note crucially that the cubic Galileon is meant only as an approximate low (or intermediate) energy effective theory, and there is no requirement that effective theories have well posed dynamics, only that their ultraviolet (UV) completions do\footnote{For works addressing precisely this issue for Galileons see \cite{Adams:2006sv,Burrage:2011cr,Keltner:2015xda,deRham:2017imi}, and for example of theories whose low energy theory is ill-posed but whose UV completion is well defined see \cite{Babichev:2017lrx}.}.

\subsection{Modified GABE}

In what follows we shall describe the numerical methods employed to solve the full Galileon interacting systems and manifest the presence of a Vainshtein mechanism. 
No assumptions are made upon the symmetries of the system.
We are using a heavily modified version of GABE to perform a $(3+1)$-dimensional finite differencing scheme to integrate the Galileon equations of motion \cite{Child:2013ria} \url{http://cosmo.kenyon.edu/gabe.html}. 
We are working in the same binary system as described in \app\ref{app:galrad} and summarized previously in \sect\ref{sec:Context}, corresponding to two sources of equal masses $M_1=M_2=M/2$ in a  circular orbit (no eccentricity).
As a first step we can consider the back-reaction of the Galileon field on the geometry (\ie on the orbits of the two objects) to be negligible.
We can then evaluate the contribution of the Galileon field within that setup and check the consistency of our assumption after the fact.
Since our numerical results will end up being in good agreement with the analytic ones performed in \cite{deRham:2012fw}, one can refer to that analysis for the consistency check where it was shown that so long as we are well inside the Vainshtein region (which is the case for the two orbiting sources), the back-reaction of the Galileon field is indeed negligible (within the level of precision we are working with).
Since we are working on a discrete grid we utilize a discretized field, that is
\begin{align}
\label{eqn:discretization}
	\pi(t,x,y,z) &\longrightarrow \pi_{ijk}(t)=\pi(t, -\rbox+i\Delta x,-\rbox+j\Delta y,-\rbox+k\Delta z),
\end{align}
where $\Delta x= \Delta y= \Delta z$ is the distance between adjacent grid points.
When taking spatial derivatives we use the standard second order center differencing stencils. 
As with most time-stepping algorithms, $\dot \pi$ is treated as its own field and is discretized similarly.

The whole essence of the Vainshtein mechanism \cite{Vainshtein:1972sx} precisely relies on the fact that the linear theory is not representative of the physics of the system, and one needs to rely on non-linear interactions even in a regime which would typically qualify as the weak-field from a standard GR point of view.
However these non-linearities are what make it so difficult to solve for the system analytically and drive the need to develop numerical methods.
Even relying on numerical techniques, the fact that one needs to deal with non-linearities in derivatives of the field makes the problem challenging to evolve numerically.
Here we discuss the equations of motion we are evolving and the steps we have undertaken to alleviate the numerical issues arising from their non-linear nature.

Varying the action, \eqn\eqref{eqn:galaction}, gives us the equations of motion for the cubic Galileon as
\begin{equation}
	\Box\pi +\frac{1}{3\Lambda^3}\left(  (\Box\pi)^{2} - (\partial_\mu\partial_\nu\pi)^2 \right) = -\frac{T}{3 \mpl}.
\end{equation}
Rescaling the problem as described in \app\ref{app:rescale} we see that
\begin{equation}
	\label{eqn:cubic_eom_prg}
	\Box_{\prg}\pi_{\prg} +\frac{\kappa}{3}\left((\Box_{{\text{pr}}}\pi_{\prg})^{2} - (\partial^{\text{pr}}_\mu\partial_\nu^{\text{pr}}\pi_{\prg})^2 \right) = -\frac{\lambda}{3} T_{\text{pr}}.
\end{equation}
We regularize the delta functions in the source as Gaussians, so in program units the source becomes,
\begin{equation}
	T_\prg = -\frac{1}{\left(2\pi\sigma\right)^{3/2}}\left(e^{-\frac12\left(\vec r_1{}^\prg(t_\prg)/\sigma\right)^2}+e^{-\frac12\left(\vec r_2{}^\prg(t_\prg)/\sigma\right)^2}\right),
\end{equation}
where $\vec r_{1,2}(t)=\{x\pm\cos(\op t),y\pm\sin(\op t),z\}$.

\subsection{Boundary \& Initial Conditions}

Beside solving for the actual dynamical equations, much of the physics relies on the appropriate boundary and initial conditions.

\paragraph{Boundary Conditions at the Origin:}

One strong advantage to using a 3D Cartesian mesh is that there is no special treatment of the origin of the coordinate system like there is in spherical or cylindrical coordinates. 
This means that we need not enforce analyticity or other such conditions at the origin to ensure a healthy evolution of our theory, as these conditions will evolve naturally out of the initial conditions and the equations of motion.

\paragraph{Absorbing Boundary Conditions:}
To allow the energy to radiate out of the system we impose absorbing boundary conditions, assuming that the radiation is leaving in spherical waves,
\begin{align}
\label{eqn:bc_wkb}
	\dot\pi = -\left(\pi/r+\partial_r \pi\right).
\end{align}
Of course these boundary conditions are strictly valid only for a Klein-Gordon field in the WKB regime, so there are corrections to this for a generic Galileon.
However, we have found that these corrections due to the Galileon are small as long as we are outside of $\rv$ (where the Klein-Gordon terms dominate) and, more importantly, we are deep in the WKB regime. 
That is $\rbox>\rv> \op^{-1}$.
When we satisfy this condition, we find that the actual amount of reflected radiation is minimal and we see little correction to the quadrupole power from $\rbox$.

\paragraph{Initial Conditions:} We set up the initial conditions by ensuring that the field is in its vacuum $\pi\equiv 0, \dot\pi\equiv 0$  when $T=0$, and evolve numerically from that initial condition.
As discussed later, Galileon models can exhibit various branches of solutions.
Setting the initial condition $\pi=0$ and $\dot \pi=0$ in the vacuum ensures that the solution we obtain is the one that connects continuously to the standard branch of the theory (see \cite{deRham:2016plk} for a discussion on how the Vainshtein mechanism may manifest itself slightly differently in other no-trivial branches).

\begin{figure}[!ht]
\centering
\includegraphics[width=0.49\textwidth]{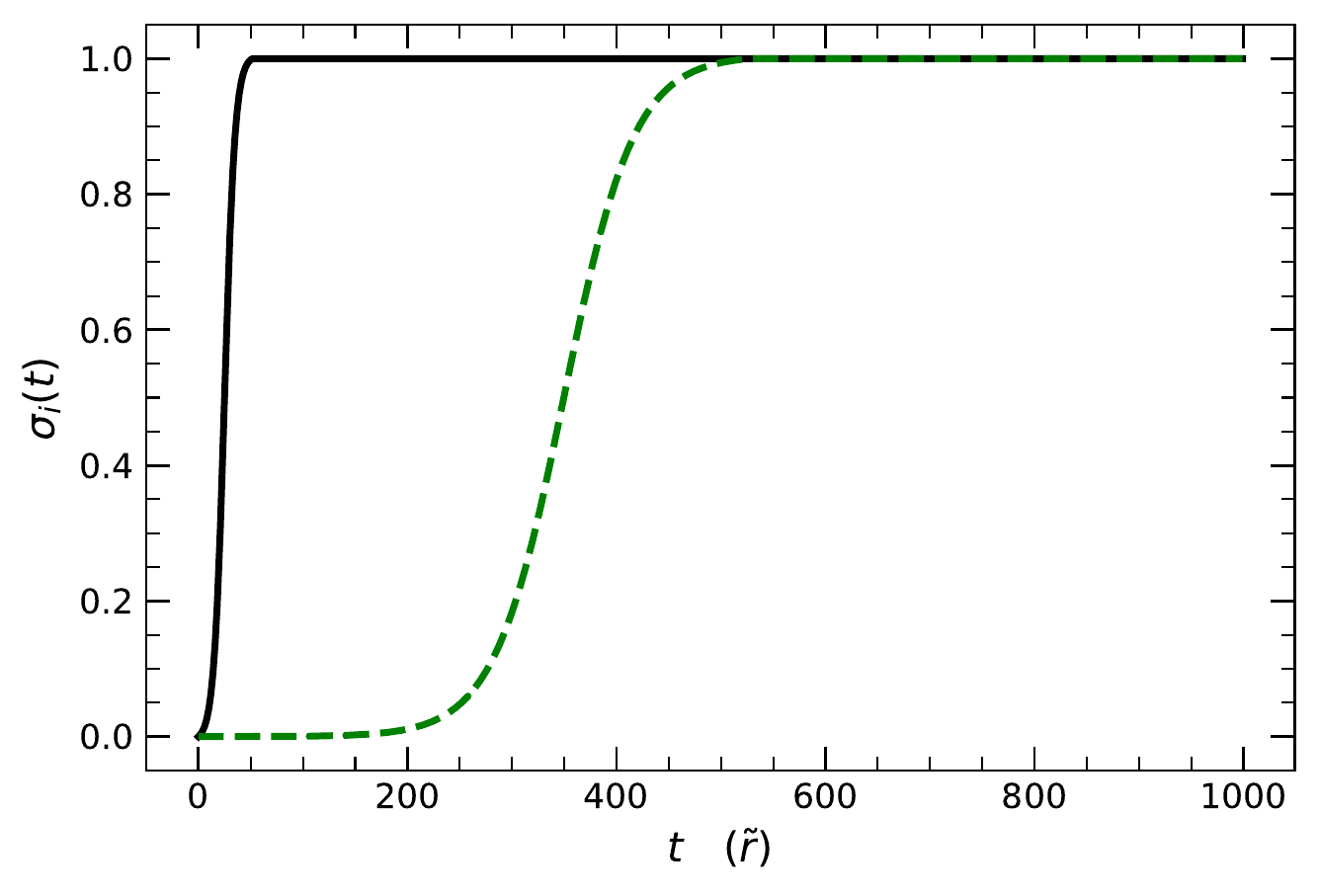}
\includegraphics[width=0.49\textwidth]{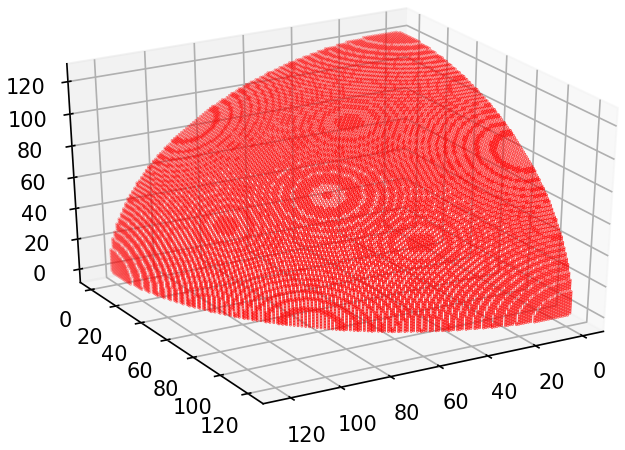}
\caption{\emph{Left Panel:} The window functions $\sigma_1(t)$ and $\sigma_3(t)$ from \eqn\eqref{eqn:window_function}, for the source term (solid line) and the non-linear terms (dashed line) for typical parameter values $t_1=25\tilde r,t_3=350\tilde r,s_1=0.1\tilde r^{-1},s_2=0.015\tilde r^{-1}$ and $o_1=o_3=0.01$.
\emph{Right Panel:} The set of points sampled in the positive octant of $S^2_r$ at a resolution of 256 points per side. }
\label{fig:window}
\label{fig:discrete_sphere}
\end{figure}

From this vacuum initial condition we slowly switch on the source.
More precisely we multiply the source by the window function depicted in the left panel of \fig\ref{fig:window} so that the effect of the source vanishes on the initial surface and the source is fully turned on after a few hundred $\tilde r$.
Tests were performed with various other shapes of window function. 
We found that as long as the process is slow enough to not shock the system there are no observable differences in the final state of the system.

When the cubic Galileon interactions are present, the simulation is more stable if we first allow the system to settle in the free-field model with the full source switched on for a few orbits, before adiabatically switching on the cubic Galileon interactions. 
This also ensures that we start with the well-known free theory solution with correct asymptotics before switching on the Galileon interactions.

While switching on the sources and the non-linear interactions energy and momentum are not conserved, so we observe significant monopole and dipole radiation during these transition phases.
This is just an artifact of how we set the system up numerically, and as we shall see later (see the left panel of \fig\ref{fig:kg_power}) those channels are then suppressed as expected once the system is fully relaxed.

With these initial conditions our equation of motion \eqn\eqref{eqn:cubic_eom_prg} becomes
\begin{equation}\label{eqn:cub_eom_window}
	\Box_{\prg}\pi_{\prg} +\sigma_3(t)\frac{\kappa}{3}\left(  (\Box_{{\text{pr}}}\pi_{\prg})^{2} - (\partial^{\text{pr}}_\mu\partial_\nu^{\text{pr}}\pi_{\prg})^2 \right) = -\frac{\lambda\sigma_1(t)}{3} T_{\text{pr}},
\end{equation}
where
\begin{align}
\label{eqn:window_function}
\sigma_i(t) &= \min(1,\max(0,\sigma_{i,0}(t))),\\
\sigma_{i,0}(t) &= \frac{\tanh(s_i(t-t_i))+\tanh(s_it_i)}{1+\tanh(s_it_i)-o_i}.
\end{align}
As previously discussed this `adiabatic' turning on of the source and subsequently the non-linear terms, helps to tame non-physical gradients growing in the simulation. Only after these terms are unity and the system has relaxed are we physically resolving the system.

\subsubsection{Numerical Constraints}

There are several aspects of this setup that challenges our ability to numerically evolve the system. 
The two foremost being the larger hierarchy of scales $\bar r\ll 1/\op \ll \rv$ and the non-linear nature of the Galileon equation of motion.

\paragraph{Hierarchy of Scales:}
The large hierarchy of scales poses a particular problem, as we need to resolve the scale $\tilde r=\bar r/2$ (see \eqn\eqref{eqn:tildr_def}) over the whole box since this is the relevant scale for the radiation.
Furthermore, only if $\op\tilde r\ll 1$ can we safely ignore relativistic corrections to the orbit and the power.
This means that numerically our hierarchy is restricted to be about an order of magnitude between scales.
The fiducial simulation has $\tilde r:1/\op:\rv:\rbox$ as approximately $1:7:50:60$.
This implies that we are not strongly in the same regime as what was assumed when performing the analytic perturbation analysis in \cite{deRham:2012fw} (see \app\ref{app:galrad}). Despite the weaker hierarchy of scales we still demonstrate the same power law dependence on $\op$ and $\rv$ as found in the perturbation theory for the cubic Galileon which we shall see later (see \figs\ref{fig:kg_power_dep}, and \ref{fig:power_omega}).

Using an adaptive mesh or other multi-grid method may help alleviate some of the tension between the hierarchy of scales especially between the length scale $\tilde r$ and $\rv$. These methods also run the risk of biasing the simulation on where relevant physics lies based on the statistics that determine the mesh resolution.

\paragraph{Courant-Friedrichs Lewy Condition:}
A necessary condition for stability of the simulations can be estimated by the Courant-Friedrichs-Lewy (CFL) condition.
The condition is
\begin{equation}
	\Delta t\left(\frac{v_x}{\Delta x}+\frac{v_y}{\Delta y}+\frac{v_z}{\Delta z}\right)<1
\end{equation}
where $v$ is the maximum magnitude of velocity of the field, $\Delta t$ the time resolution, and $\Delta x=\Delta y=\Delta z$ is the spatial resolution of the simulation (for our simulations each direction is treated identically).
About a spherically symmetric source the velocity can go up to $v\approx 1.3$, but this velocity can be much higher during the adiabatic turning on of the non-linear terms.
This is an unphysical artifact of the way we set up the numerical system, but one that needs to be addressed in order for the full time evolution of the fields to be stable.
To satisfy the CFL condition we typically evolve with $20 \Delta t<\Delta x$.

\subsubsection{Numerical Power}
To evaluate the power on the grid we compute the radial flux of the scalar field $\pi$ over some radius $\rbox$ larger than $\rv$, so that we are outside the strong coupling regime.
The stress energy associated with $\pi$ is
\begin{eqnarray}
	t_{\mu\nu}^\pi &=& \frac32 \left(\partial_\mu\pi\partial_\nu\pi-\frac12\eta_{\mu\nu}(\partial\pi)^2\right)	+\frac1{2\Lambda^3}\partial_\mu\pi\partial_\nu\pi\Box\pi\\
	&-&\frac1{4\Lambda^3}\left(\partial_\mu(\partial\pi)^2\partial_\nu\pi+\partial_\nu(\partial\pi)^2\partial_\mu\pi - \eta_{\mu\nu}(\partial\pi)^2\Box\pi\right)\,.\nn
\end{eqnarray}
The radial flux is simply $t_{0r}^\pi$. For simplicity we perturb the field $\pi$ around the static background and the flux reduces to
\begin{align}
	t_{0r}^{\pi}=\frac32\left(1+\frac{4}{3\Lambda^{3}}\frac{E}{r}\right)\partial_{t}\pi\partial_{r}\pi,
\end{align}
where $E$ is defined as $E(r)\hat r = \vec\nabla\pi_0$ and is explicitly given by \eqn\eqref{eqn:cub_bkg}.
The integral of this over the sphere of radius $r$ gives the radiated power,
\begin{align}
	\label{eqn:power_int}
	P&=\frac{3r^{2}}{2}\left(1+\frac{4}{3\Lambda^3}\frac{E}{r}\right)\int \diff\Omega\,\partial_{t}\pi\partial_{r}\pi\,,
\end{align}
where $E$ is defined in \eqn\eqref{eqn:cub_bkg}.
The dependence of the calculated power on this term decreases as $\rbox\gg\rv$. 
However, for the small hierarchies in which we are working in this correction factor is necessary.

Upon discretization (\eqn\eqref{eqn:discretization}), numerically evaluating this integral corresponds to summing over discrete points on the boundary of the 2-sphere of radius $\rbox$ centered on the origin ($S^2_r$). 
The integral \eqref{eqn:power_int} becomes the sum
\begin{align}
	P^{\text{tot}}_{\text{comp}}(t,r)=\frac{3}{2}\left(\frac{4\pi r^2}{N_s}\right)\left(1+\frac{4}{\Lambda^3}\frac{E}{r}\right)\sum_{\{i,j,k\}\subset S^2_r}\partial_{t}\pi_{ijk}(t)\partial_{r}\pi_{ijk}(t)\,,
\end{align}
where the sum is over the set of $N_s$ points on the 2-sphere $S^2_r$. 
The discretized field $\pi_{ijk}(t)$ is defined in \eqn\eqref{eqn:discretization}.
Technically the angle element $\diff\Omega$ should be calculated individually for each point on the sphere; however for large enough resolution the difference in solid angle between points is small. 
Thus the solid angle can safely be approximated by its average over the sphere.
See the right panel of \fig\ref{fig:discrete_sphere} for an example of what the subset of the sphere $S^2_r$ we sample looks like for a simulation of resolution $256^3$ and $\rbox=60\tilde r$. 
A plot of the instantaneous power is given in the left panel of \fig\ref{fig:power_not_avg}.

\subsection{Free-field}
Since we are numerically constrained to have a poor hierarchy between $\rbox$ and $\op^{-1}$ (at best an order of magnitude), we first perform several simulations with just the free Klein Gordon scalar field (\ie just the kinetic term without any Galileon Vainshtein screening, $\kappa=0$). 
This ensures that our boundary conditions (\eqn\eqref{eqn:bc_wkb}) are still valid even though we are not  strongly in the WKB regime.

\begin{figure}[!ht]
\centering
\includegraphics[width=0.49\textwidth]{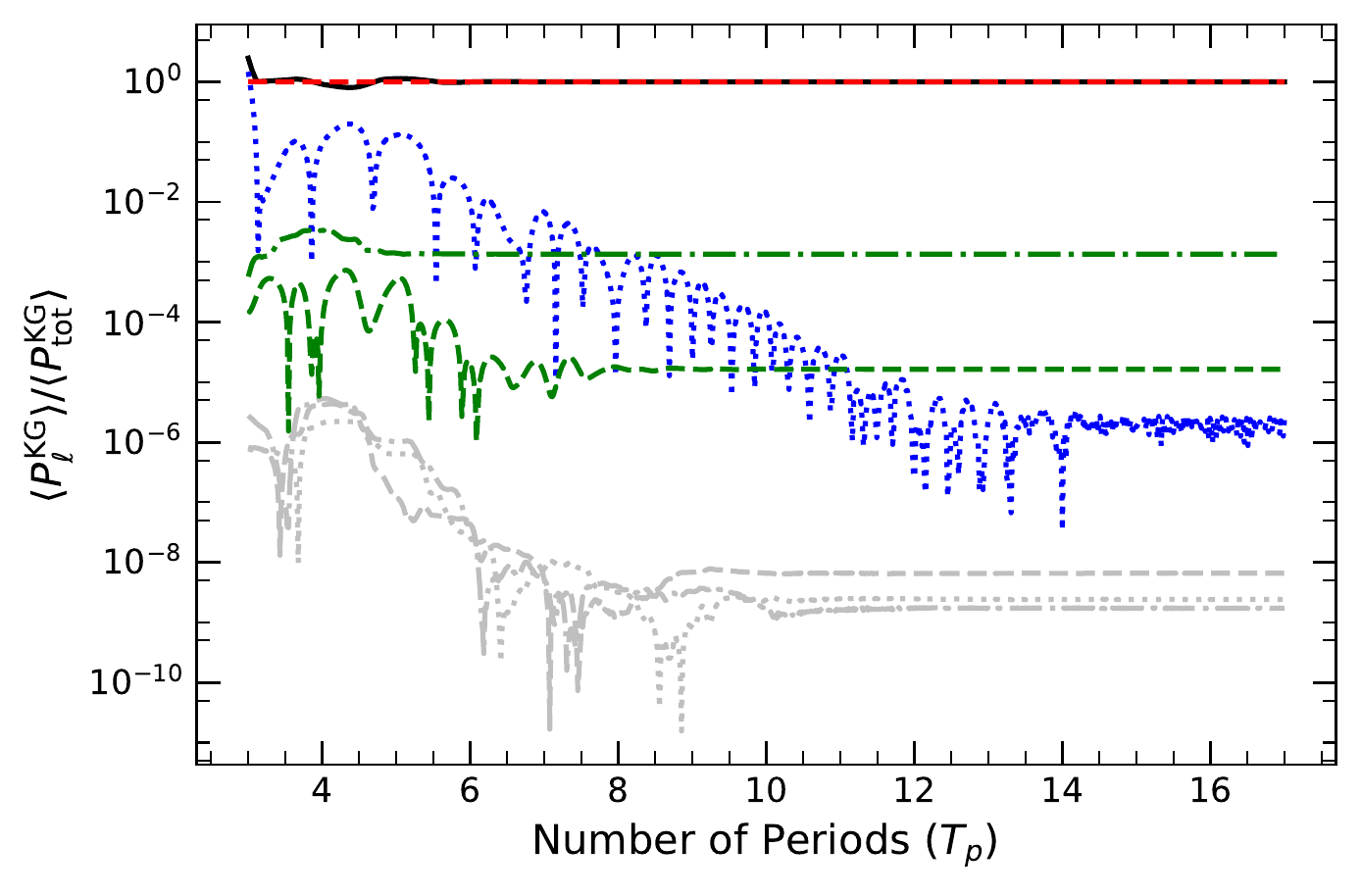}
\includegraphics[width=0.49\textwidth]{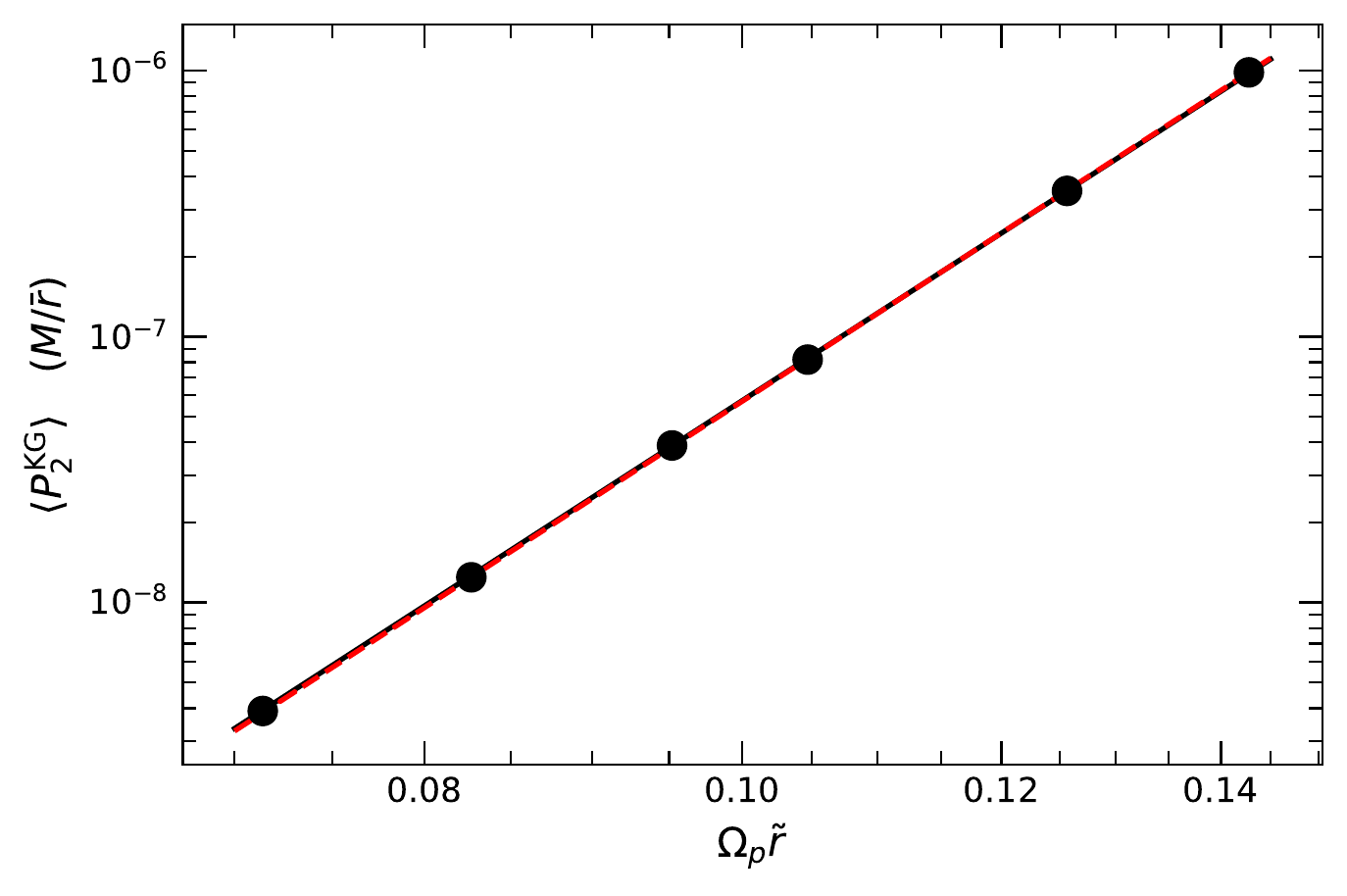}
\caption{\emph{Left Panel:} Free field ($\kappa=0$)  time averaged power in each multipole divided by the total power at late time for the simulations  with parameters $\rbox = 60\tilde r$, and $\op\tilde r = \pi/22$.
Total Power: solid black, Monopole: dotted blue, Dipole: dotted gray, Quadrupole: dashed red, $l=3$: dashed gray, $l=4$: dot-dashed green, $l=5$: dot-dashed gray, $l=6$: dashed green.
\emph{Right Panel:} Late time, time averaged quadrupole power for simulations of the free Klein Gordon field ($\kappa=0$) with $\op \in \{\pi/22,\pi/25,\pi/30,\pi/33,\pi/38,\pi/44\}$ and $\rbox=60\bar r$.
The analytic power \eqn\eqref{eqn:kgp2_pr} is shown as the solid black line where the analytic expectation ($P_{\text{tot}}^\text{th}\propto \omega^{8}$) is the dashed red line.}
\label{fig:kg_power}
\label{fig:kg_power_dep}
\end{figure}

We find that even down to a hierarchy of $\rbox=5\times \op^{-1}$, the corrections to the power coming from the higher multipoles come in at less than 0.5\%.
The left panel of \fig\ref{fig:kg_power} shows the power in each multipole over the duration of the simulation.
We consider the simulation fully relaxed from the adiabatic turning on of the source just after $t= 8T_p$.

Furthermore, we find that the numerically calculated power agrees with the analytically calculated power (\eqn\eqref{eqn:kgp2_pr}) to within 1\% agreement.
The agreement between \eqn\eqref{eqn:kgp2_pr} and the simulated quadrupole power can be seen in the right panel of \fig\ref{fig:kg_power_dep}, where we show the quadrupole power for simulations with $\op \tilde r\in \{\pi/22,\pi/25,\pi/30,\pi/33,\pi/38,\pi/44\}$ and the analytic prediction (dashed black line).

The strong agreement between the numeric and the analytic  power for the Klein-Gordon field, despite not strongly being in the WKB regime, allows us to move on with confidence to working with the cubic Galileon numerically.

\subsection{Cubic Galileon}
\begin{figure}[!ht]
\centering
\includegraphics[width=0.49\textwidth]{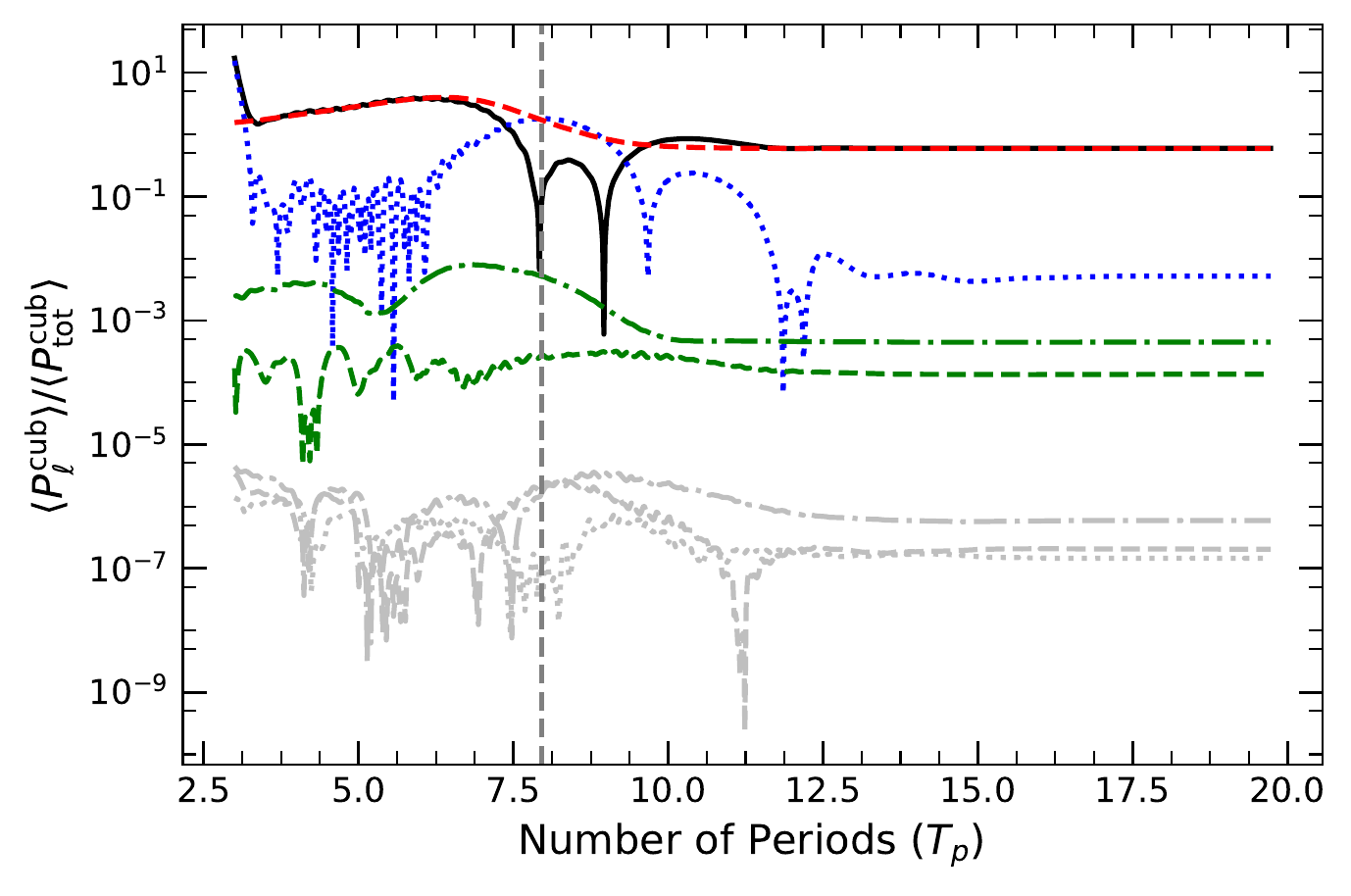}
\includegraphics[width=0.49\textwidth]{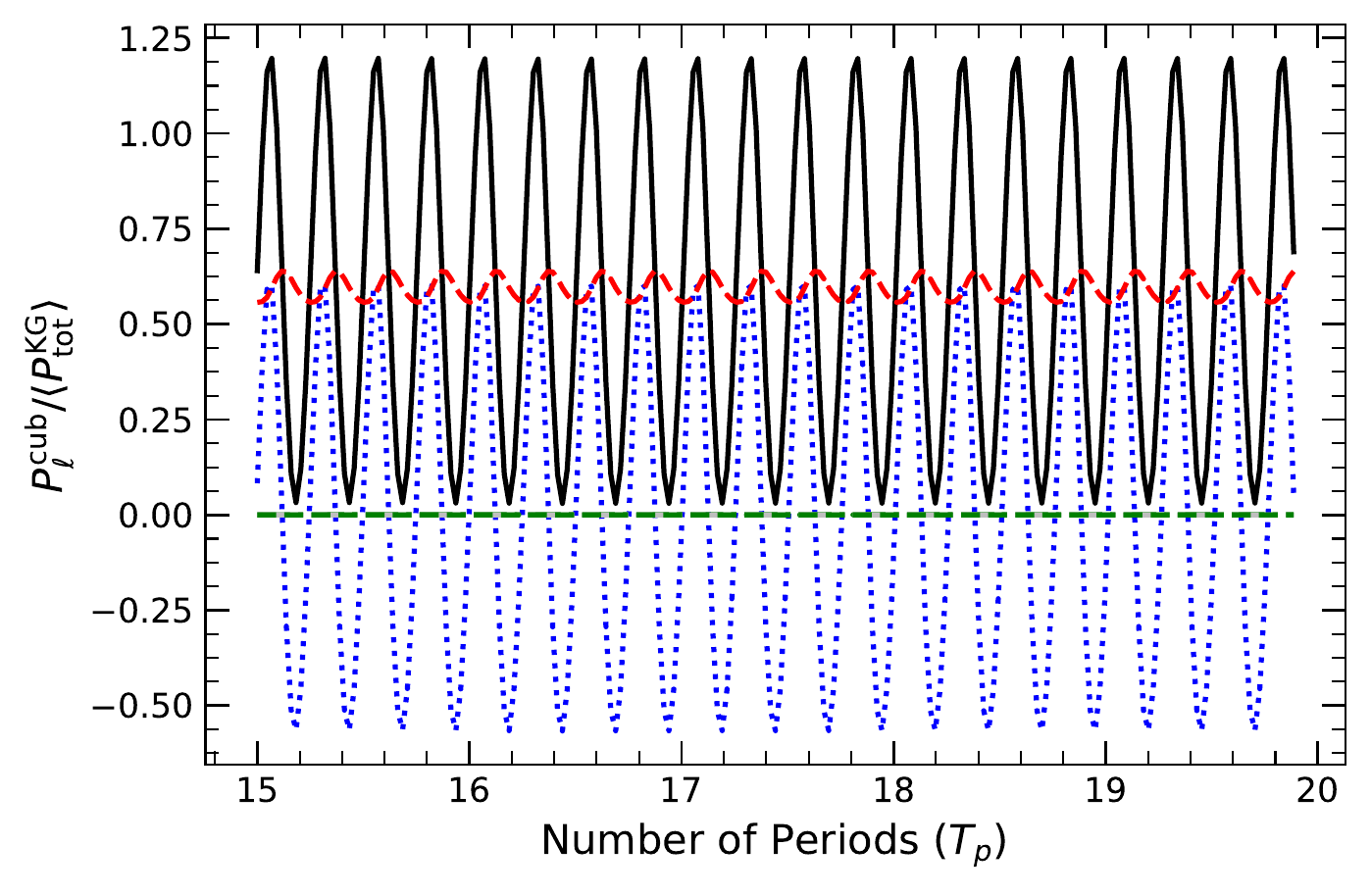}
\caption{Cubic Galileon time averaged power (left) and instantaneous power (right) in each multipole divided by the total power at late time for the fiducial parameters $\rbox = 60\tilde r,\rv = 50 \tilde r$, and $\op\tilde r = \pi/22$.
Total Power: solid black, Monopole: dotted blue, Dipole: dotted gray, Quadrupole: dashed red, $l=3$: dashed gray, $l=4$: dot-dashed green, $l=5$: dot-dashed gray, $l=6$: dashed green. The vertical dashed gray line is at $t_3=350\tilde r.$, when the non-linear terms are turning on.}
\label{fig:fiducial_power}
\label{fig:power_not_avg}
\end{figure}

\begin{figure}[!ht]
\centering
\includegraphics[width=0.9\textwidth]{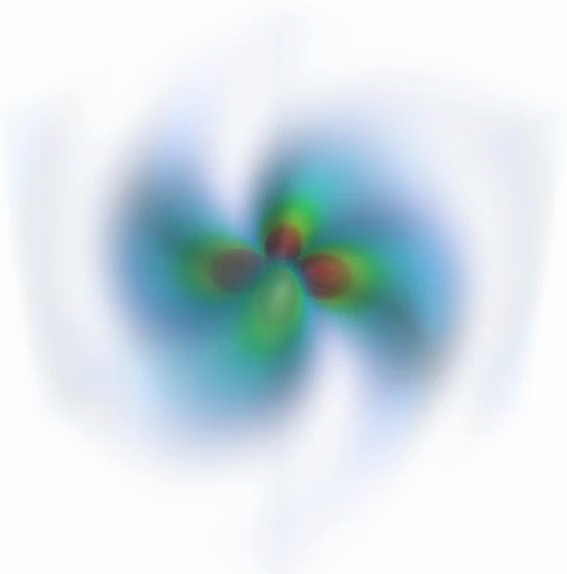}
\caption{Energy density of the cubic Galileon field after the simulation has relaxed $t=22T_p$ for $\rbox = 60\tilde r,\rv = 50 \tilde r$, and $\op\tilde r = \pi/22$.
Red is higher energy density and blue lower.
\label{fig:energy_den}
}
\end{figure}

We now turn on the cubic interaction $\kappa> 0$, keeping the same standard kinetic term as in the previous section.
The fiducial simulation is performed at a resolution of $384^2$ with the length along one side of the simulation $L=120\bar r$, $\op\tilde r=\pi/22$, $\rv=50\tilde r$, corresponding to dimensionless numerical parameters $\lambda\sim 0.7$ and $\kappa \sim 10^7$.
The simulation ran for $t_f=1000\tilde r$, which included $7T_p\simeq 300\tilde r$ of time where the simulation was fully relaxed.
The results of the simulation are shown in \fig\ref{fig:fiducial_power}, where the time averaged power is given in units of the total final power (left panel).
The simulation has fully relaxed at about $15T_p\simeq650\tilde r$.
The instantaneous power is shown after the simulation has relaxed in the right panel of \fig\ref{fig:power_not_avg}.
As predicted by perturbative analysis the quadrupole is the dominant mode, containing more than 99\% of the total radiated power. 
This can be seen visually in the plot of the energy density \fig\ref{fig:energy_den}.
Oddly the monopole is the next dominant mode.
This is likely due to a poor hierarchy. 
That is, both $\rv\not\gg\op^{-1}$ and $\rbox\not\gg\op^{-1}$, implying that we are not computing the power deeply in the linear WKB regime.

\begin{figure}[!ht]
\centering
\includegraphics[width=0.49\textwidth]{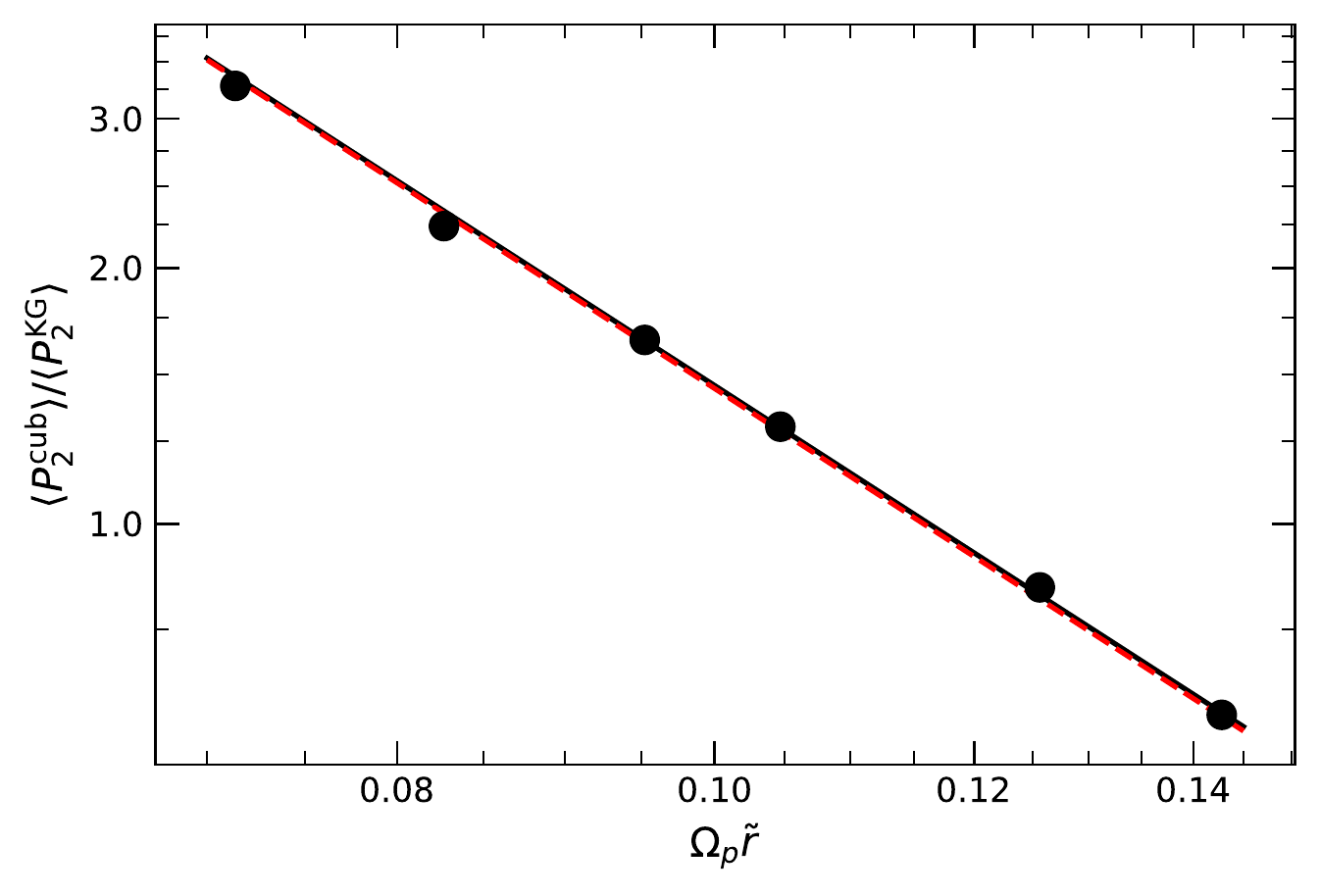}
\includegraphics[width=0.49\textwidth]{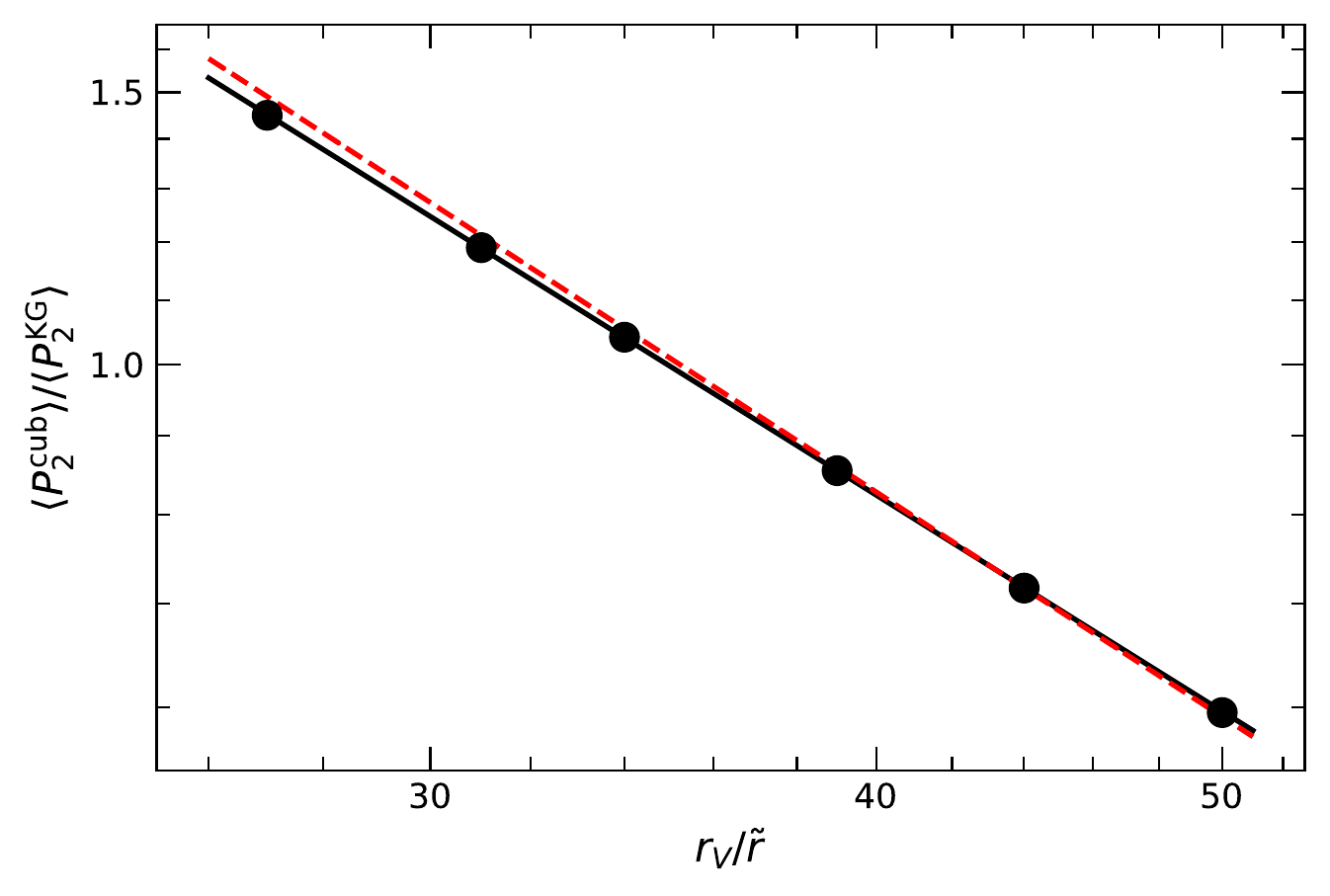}
\caption{Cubic Galileon late time, time-averaged quadrupole power from simulation divided by the Klein-Gordon power. 
The left panel shows simulations with $\rv=50 \bar r$ and $\op \in \{\pi/22,\pi/25,\pi/30,\pi/33,\pi/38,\pi/44\}$.
The best fit for $P_{2}^\text{cub}/P_2^\text{KG}\ \propto \ \omega^{-2.49}$ is the solid black line where the best fit with the analytic scaling ($P_{2}^\text{theory}/P_2^\text{KG}\ \propto \  \omega^{-5/2}$) is the dashed red line.
The right panel shows simulations with $\op\tilde r=\pi/22$ and $\rv/\tilde r \in \{50, 44, 39, 34, 31, 27\}$.
The best fit for $P_{2}^\text{cub}/P_2^\text{KG}\ \propto \ \rv^{-1.44}$ is the solid black line where the best fit with the analytic scaling ($P_{2}^\text{theory}/P_2^\text{KG}\ \propto \ \rv^{-3/2}$) is the dashed red line.}
\label{fig:power_omega}
\label{fig:power_rv}
\end{figure}

We show ten other simulations to probe the dependence on the quadrupole power on both $\op$ and $\rv$, depicted in \fig\ref{fig:power_omega}.

\subsubsection{Dependence on Orbital Period}
As $\op$ shrinks we find that the quadrupole mode remains the dominant mode, always containing more than 98\% of the power. 
The monopole mode grows, but always stays at less than 2\% of the total power (see \tbl\ref{table1}).
Increasing $\op$ is constrained by requiring minimal relativistic corrections to the orbit and power.
Despite these constraints, the computed quadrupole power dependence on $\op$ gives us
\begin{eqnarray}
\left.\frac{P_2^\text{cub}}{P_2^\text{KG}}\right|_\text{numeric}\ \propto\  \op^{-2.49}\quad \text{while}\quad
\left.\frac{P_2^\text{cub}}{P_2^\text{KG}}\right|_\text{analytic}\ \propto\  \op^{-5/2}\,.
\end{eqnarray}
We therefore see a remarkable agreement with the expected analytic dependence derived in \eqn\eqref{eqn:cu_power_rat}.

\subsubsection{Dependence on Vainshtein Radius}
As $\rv$ shrinks we weaken the hierarchy $\op^{-1}\ll\rv$. 
We find that we can go down to a relative hierarchy of $\rv\approx 3\op^{-1}$ and maintain more than 98\% of the power in the quadrupole mode.
Similarly, increasing $\rv$ is constrained by being able to resolve all scales of the problem.
We find that the quadrupole mode remains the dominant mode, always containing more than 98\%.
Despite these constraints the computed quadrupole power dependence on $\rv$ gives us
\begin{eqnarray}
\left.\frac{P_2^\text{cub}}{P_2^\text{KG}}\right|_\text{numeric}\ \propto \ \rv^{-1.44}\quad \text{while}\quad
\left.\frac{P_2^\text{cub}}{P_2^\text{KG}}\right|_\text{analytic}\ \propto \ \rv^{-3/2}\,,
\end{eqnarray}
which is again in good agreement with the expected dependence of \eqn\eqref{eqn:cu_power_rat}.

\begin{table}[!ht]
\centering
\begin{tabular}{cc|clcc}
$\op\tilde r$ & $\rv/\tilde r$ & $P^\text{cub}_\text{tot}/P^\text{KG}_\text{tot}$ & $P^\text{cub}_\text{0}/P^\text{KG}_\text{tot}$ & $P^\text{cub}_\text{1}/P^\text{KG}_\text{tot}$ & $P^\text{cub}_\text{2}/P^\text{KG}_\text{tot}$ \\
\hline
$\pi/22$ & 27 &  1.5 & \ \ 0.01 &  3e-08 &  1.4 \\
$\pi/22$ & 31 &  1.2 & \ \ 0.01 &  4e-08 &  1.2 \\
$\pi/22$ & 34 &  1.0 & \ \ 0.01 &  4e-08 &  1.0 \\
$\pi/22$ & 39 &  0.9 & \ \ 0.007 &  5e-08 &  0.9 \\
$\pi/22$ & 44 &  0.7 & \ \ 0.006 &  8e-08 &  0.7 \\
$\pi/22$ & 50 &  0.6 & \ \ 0.005 &  1e-07 &  0.6 \\
$\pi/25$ & 50 &  0.9 & \ \ 0.01 &  6e-08 &  0.8 \\
$\pi/30$ & 50 &  1.3 & \ \ 0.03 &  1e-08 &  1.3 \\
$\pi/33$ & 50 &  1.7 & \ \ 0.05 &  1e-07 &  1.6 \\
\end{tabular}
\caption{Period averaged total power and multipole power (monopole, dipole and quadrupole) radiated by the cubic Galileon divided by the power radiated by a Klein-Gordon field for the same system. Note that the monopole power is suppressed by a factor of 100 where the dipole a factor of $10^7$ with respect to the total power. In all cases the quadrupole is the dominant power.}
\label{table1}
\end{table}

\section{Conclusion}
\label{sec:conc}

We have successfully performed full four dimensional simulations of a cubic Galileon coupled to a binary system on Keplerian orbits, and computed the resulting radiated scalar gravitational power.
Our numerical results exhibit a power law dependence on the parameters $\op$ and the Vainshtein radius $\rv$, relative to the GR result of the form
\begin{equation}
	\left.\frac{P^\text{cubic}_2}{P^\text{GR}_2}\right|_{\rm numeric} \propto (\op)^{-2.49 } (\rv)^{-1.44}.
\end{equation}
This is in very good argument with the scaling predicted by the perturbative analytic results derived in \cite{deRham:2012fw} (\ie \eqn\eqref{eqn:cu_power_rat}) assuming a strong hierarchy between $\bar r\ll\op^{-1}\ll\rv$. 
Even though the numerical simulations are performed with a relatively mild hierarchy, the consistency of the two lends strong support to the validity of the scaling implied by the analytic approximations performed in \cite{deRham:2012fw} over a much wider range of parameter space than is accessible numerically. 
At present the large hierarchies relevant to realistic binary pulsar systems for the cosmologically motivated choice of the scale $\Lambda$ are beyond the reach of our simulations, purely for reasons of storage and computational time. 
Nevertheless the agreement between the numerical and analytic results in the regime in which both are expected to be valid allows us to be confident in the analytic scaling.

An important future step will be to apply these simulations to the quartic and quintic Galileon theories. 
These theories are of more direct relevance to hard massive gravity theories. 
In this case the approximations performed in \cite{deRham:2012fg} simply broke down, and so at present we do not have an analytic result to check or to make predictions. 
Analogous numerical simulations will fill this gap, and it is plausible that an improved approximate analytic treatment may have a better regime of validity. We will leave these considerations to future work.\\

\noindent{\textbf{Acknowledgments:}}
We would like to thank Matthew Johnson, Luis Lehner, Andrew Matas, James Mertens, Leo Stein, and Jun Zhang for useful discussions.
CdR and AJT would like to thank the Perimeter Institute for Theoretical Physics for hosting them during the final part of this work.
The work of CdR and AJT is supported by an STFC grant ST/P000762/1. CdR thanks the Royal Society for support at ICL through a Wolfson Research Merit Award. CdR is supported by the European Union's Horizon 2020 Research Council grant 724659 MassiveCosmo ERC-2016-COG and by a Simons Foundation award ID 555326 under the Simons Foundation's Origins of the Universe initiative, `\textit{Cosmology Beyond Einstein's Theory}'. AJT thanks the Royal Society for support at ICL through a Wolfson Research Merit Award. JTG is supported by the National Science Foundation, Grant No. PHY-1719652.  Numerical simulations were performed on equipment provided by the Kenyon Department of Physics and the National Science Foundation. This work also made use of the High Performance Computing Resource in the Core Facility for Advanced Research Computing at Case Western Reserve University.


\begin{appendix}

\section{Cubic Galileon Radiation}

\label{app:galrad}

For convenience, in this appendix we shall reproduce the analytic calculation of the power emitted in scalar waves for a free theory and a cubic Galileon exhibiting the Vainshtein mechanism as derived in \cite{deRham:2012fw,deRham:2012fg}. 
We therefore consider a cubic Galileon with  conformal coupling to matter
\begin{align}
	\label{eqn:galactionapp}
	S = \int \diff{^4 x}\left(-\frac34 (\partial\pi)^2-\frac1{4\Lambda^3} (\partial\pi)^2 \Box \pi + \frac{1}{2\mpl}\pi T \right)\,.
\end{align}
In the limit $\Lambda\to \infty$ we recover a free Klein-Gordon scalar field coupled to an external source. In that limit (corresponding to the large masses in the context of massive gravity $\Lambda=(\mpl m^2)^{1/3}$), there is no Vainshtein screening and since the coupling to external source arises at the same $\mpl$ scale as in GR, the power emitted by these scalar waves would be of the same order of magnitude as the gravitational power emitted in GR.
Actually even higher as we would expect monopole and dipole radiation which are typically suppressed by fewer powers of angular speed as compared with the quadrupole. 
In the non-relativistic limit the monopole and dipole vanish by energy and momentum conservation, but not once relativistic corrections are considered.

In the rest of this appendix we recall how to derive the analytic expressions for the period averaged power radiated via the effective action method, and specifically compute the leading contribution given by the quadrupole power  for two equal mass objects in circular orbits (zero eccentricity).
Computing this for non-zero eccentricity is slightly more complicated, as suggested in \cite{deRham:2012fw}. Since the numerical simulations focus on non-eccentric orbits, this case will be sufficient to compare between the perturbative analytic results and the numerical ones.

\subsection{Center of Mass Split}

Since we are interested in equal mass binary systems, the total stress-energy tensor $T\mupn$ is well approximated by the sum of two delta functions,
\begin{align}
	T\mupn= -M\sum_{i=1,2} \delta^{(3)}(\vec x -\vec x_i(t))\delta_0^\mu\delta_\nu^0\,.
\end{align}
The key ingredient of the analytic approach is to split the source into the static and spherically symmetric center of mass contribution and departures from it,
\begin{equation}
T\mupn = {T_0}\mupn+\delta T\mupn\,,
\end{equation}
with
\begin{equation}
{T_0}\mupn = -2M\delta^{(3)}(\vec x)\delta_0^\mu\delta_\nu^0\,.
\end{equation}
${T_0}\mupn$ leads to a static and spherically symmetric field background $\pi_0$ and in all generality,  the full exact solution can always be split as
\begin{equation}
\pi(t, \vec x) = \pi_0(r)+\sqrt{2/3}\phi(t,\vec x)\,.
\end{equation}
At this level those splits are purely mathematical and rely on no assumption. 
The only assumptions that will be performed in the analytic derivation is a large hierarchy between the different scales involved, and as a consequence $\phi$ can be treated linearly. 
Whether or not an actual hierarchy between the orbit size $\bar r$, the inverse frequency scale $\op^{-1}$, and the Vainshtein radius $\rv$ (\eqn\eqref{eqn:rv}), is present, depends on the specific scales chosen but this hierarchy is realized for the binary systems we have in mind.
Whether or not $\phi$ can be treated linearly is an assumption which can be checked after the fact as seen in \app~E(for the cubic) and \sect~3.2 (for the quartic/quintic) of \cite{deRham:2012fg}. 
Once the appropriate hierarchy of scales has been considered, fluctuations on top of the static and spherically symmetric background are indeed small and can be treated linearly.

The center of mass source $T_0$ leads to a field profile that satisfies
\begin{align}
	\label{eqn:bkg_form}
	\frac{E(r)}{r} + \frac{2}{3\Lambda^3}\left(\frac{E(r)}{r}\right)^2  = \frac{1}{12\pi r^3}\frac{M}{\mpl},
\end{align}
where $E(r)\hat r = \vec\nabla\pi_0$. There are two branches of solutions for $E$, and we focus on the `normal' branch which behaves as a free field (leading to a well-known Newton square law $E \ \propto\ 1/r^2$) for $r\gg \rv$.

The quadratic action for the perturbation $\phi$ is then
\begin{align}
	\label{eqn:pertquadact}
	\mathcal S_\phi &= \int \diff{^4 x}\left(-\frac12 Z^{\mu\nu}\partial_\mu\phi\partial_\nu\phi+\frac{\phi\delta T}{\sqrt{6}\mpl}\right)\,,
\end{align}
where $Z^{\mu\nu}$ is diagonal with non-vanishing components given by
\begin{align}
	\label{eqn:eff_comp}
	Z^{tt}(r) &= - \left[1+\frac{2 }{3\Lambda^3}\left(2\frac{E(r)}{r}+ E'(r)\right)\right] \, ,\\
	Z^{rr}(r) &= 1+\frac{4}{3\Lambda^3}\frac{E(r)}{r} \, , \\
	Z^{\Omega\Omega}(r) &= 1+\frac{2}{3\Lambda^3}\left(\frac{E(r)}{r}+ E'(r)\right)\,.
\end{align}
We define the modified d'Alembertian operator $\tilde \Box$, as
\begin{align}
	\label{eqn:pertbox}
	\tilde \Box \phi = \p_\mu \left(Z^{\mu\nu}\p_\nu\right)\phi=Z^{tt}(r)\ddot \phi + \frac{1}{r^2}\frac{\partial}{\partial r}\left(r^2 Z^{rr}(r)\frac{\partial}{\partial r}\phi\right)+Z^{\Omega\Omega}(r)\nabla_\Omega^2\phi\,,
\end{align}
and the mode functions for this operator have the form
\begin{align}
	\label{eqn:permodefunc}
	\phi_{lm\omega}(t,r,\theta,\phi) = u_{l\omega}(r)Y_{lm}(\theta,\phi)e^{-i \omega t}.
\end{align}
Imposing periodicity $T_\mathrm{P}$ on the mode functions forces  $\omega$ to be discretized, $\omega\to n\op$ for integer $n$.

\subsection{Radiation from the Effective Action}
\label{sec:eff_act_pow}
Following \cite{Goldberger:2004jt,Chu:2012kz,deRham:2012fw,deRham:2012fg} we compute the power radiated by a field using  the effective action technique.
The effective action is defined by integrating out perturbatively the scalar field from \eqn\eqref{eqn:pertquadact}.
At leading order we can express $\phi$ in terms of the Feynman propagator
\begin{align}
\label{eqn:phiFeyn}
	\phi_\mathrm{F}(x) = \frac{i}{\sqrt 6\mpl}\int \diff{^4y} G_{\mathrm F}(x,y)\delta T(y),
\end{align}
where the propagator is defined as
\begin{align}
	\label{eqn:prop_def}
	\tilde\Box G_{\mathrm F}(x,y) = i\delta^4(x-y),
\end{align}
and the modified d'Alembertian operator is that given in \eqn\eqref{eqn:pertbox}. Once we have the solution for $\phi_\mathrm{F}(x)$ in \eqn\eqref{eqn:phiFeyn}, we can compute the amplitude of the non-linear operators that enter \eqn\eqref{eqn:pertquadact} and confirm that they are indeed negligible within the regime we are working in.
We write down the propagator in terms of the Wightman functions
\begin{align}
	\label{eqn:prop_wight}
	G_{\mathrm{F}}(x,y) = \Theta(x_t - y_t)W^+(x,y)+\Theta(y_t-x_t)W^-(x,y)\,,
\end{align}
where the Wightman functions are defined as
\begin{align}
	\label{eqn:wight_def}
	W^\pm(x,y) = \sum_{lm}\int_0^\infty \diff{\omega} \phi_{lm\omega}(\pm x_t,\vec x)\phi_{lm\omega}^*(\pm y_t,\vec y)\,,
\end{align}
and the $\phi_{lm\omega}$ are the complete set of mode functions defined in \eqn\eqref{eqn:permodefunc}.
Thus integrating out $\phi$ from \eqn\eqref{eqn:pertquadact} yields the effective action
\begin{align}
	S_\mathrm{eff} = \frac{i}{12\mpl^2}\int \diff{^4 x}\diff{^4 y}\delta T(x)G_\mathrm{F}\delta T(y).
\end{align}
As pointed out in \cite{Goldberger:2004jt,deRham:2012fw} the time averaged power in the system is
\begin{align}
	P = \int_0^\infty \diff{\omega} \omega f(\omega),
\end{align}
where $f(\omega)$ is related to the effective action (integrated over one period) by
\begin{align}
	\frac{2 \mathrm{Im} (S_\mathrm{eff})}{T_\mathrm{P}}=\int_0^\infty \diff{\omega} f(\omega).
\end{align}
Defining the moments
\begin{align}
	\label{eqn:gen_moments}
	M_{lmn}=\frac1{T_\mathrm{P}}\int_0^{T_\mathrm{P}}\diff t\int \diff{^3x} \phi_{lmn}(x,t) \delta T
\end{align}
and solving for $f(\omega)$ yields
\begin{align}
	f(\omega)=\frac{\pi}{3\mpl^2}\sum_{n=0}^\infty\sum_{l,m}|M_{lmn}|^2\delta(\omega - n\op).
\end{align}
Thus the period averaged power is
\begin{align}
	\label{eqn:}
	P = \frac{\pi}{3\mpl^2}\sum_{n=0}^\infty\sum_{lm}n\op|M_{lmn}|^2.
\end{align}
Consequently the power in a given mode $l$ is
\begin{align}
	P_{l}&=\frac{\pi}{3\mpl^2}\sum_{n=0}^\infty\sum_{m}n\op|M_{lmn}|^2.
\end{align}

We now restrict ourselves to circular orbits in the $\theta=\pi/2$ plane with equal mass objects. That is,
\begin{equation}
	\delta T = M \left[\delta^3(x)-\frac 12\left(\delta^3(\vec r -\vec r_1)+\delta^3(\vec r -\vec r_2)\right)\right]
\end{equation}
where $r_{1,2}=\bar r/2$, $\theta_{1,2}=\pi/2$, and $\phi_{1,2}=\op t+\pi\delta_{i,2}$.
Combining this and \eqn\eqref{eqn:permodefunc} into \eqn\eqref{eqn:gen_moments} gives us
\begin{align}
\label{eqn:red_moment}
 M_{lmn}
 &= M\left[u_{ln}(0)Y_{lm}(0,0)\delta_{n,0} -\frac{(1+(-1)^m)}{2} u_{ln}(\bar r/2)Y_{lm}(\pi/2,0)\delta_{n,m}\right].
\end{align}
We note that since there is a leading $n$ in the expression for the power, the first term of \eqn\eqref{eqn:red_moment} will never contribute to the power.
Thus we rewrite the power in a given mode $l$ as
\begin{align}
	\label{eqn:pl_reduced}
	P_{l}&=\frac{\pi \op M^2 }{6\mpl^2}\sum_{m=0}^l m(1+(-1)^m) u^2_{lm}(\bar r/2)|Y_{lm}(\pi/2,0)|^2.
\end{align}
As expected, the power radiated in the monopole mode is zero (because $l=0$ constrains $m$ to vanish).
As in GR, the monopole being zero can be understood as a consequence of conservation of energy.

Further, the power in the dipole mode is also zero because for $l=1, m=0$, the leading $m$ kills the power, and for $l=1,m=1$, the term $\left(1+(-1)^m\right)$ will be zero.
This is also understood (as in GR) as a consequence of conservation of momentum.

This means that the first non-zero multipole will be the quadrupole. Further since $m>0$ and must be even in order that $\left(1+(-1)^m\right)$ is non-zero, we know that the $m=2$ term is the only contributing term to the quadrupole power.

\subsection{Free-field (Klein-Gordon) Radiation}
For the free field ($\Lambda\to \infty$), the mode functions are the standard Klein-Gordon ones given by
\begin{equation}
\label{eqn:kgmodefull}
	\phi_{lmn}(r,\theta\phi,t) = \sqrt{\frac{n\op}{\pi}}j_l(n r\op)Y_{lm(\theta,\phi) }e^{-i \op t}\,.
\end{equation}
In realistic systems (like the Hulse-Taylor pulsar and even more so for more relativistic systems like neutron star or black hole mergers) there is a strong hierarchy of scales $\bar r\ll\op^{-1}\ll \rv$.
Further, since we are assuming a Keplerian orbit and are ignoring relativistic corrections to the orbit we are also assuming $\op\bar r\ll 1$. In this limit the radial part of the mode function is
\begin{equation}
\label{eqn:kgrmode}
	u_{ln} = \sqrt{\frac{\op}{2}}\frac{ (n \op r/2)^l }{\Gamma\left(\frac32+l\right)}.
\end{equation}
Note that this is suppressed for large $l$ and thus the dominant multipole in the power will be the quadrupole.
Using the radial mode function (\eqn\eqref{eqn:kgrmode}) to compute the power radiated in the quadrupole mode (\eqn\eqref{eqn:pl_reduced}), we find the standard result
\begin{equation}
	\label{eqn:kgp2_mplapp}
	P_2^\text{KG} =\frac{1}{45} \frac{M^2}{8\pi \mpl^2} ( \op \bar r)^4\op^2.
\end{equation}
For reference this only differs from the Peter-Mathews result (quadrupole power radiated in GR by a binary system) of \cite{Peters:1963aa}
\begin{equation}
P_\text{Peter-Mathews} =\frac{2}{5} \frac{M^2}{8\pi\mpl^2} (\op \bar r)^4	\op^2
\end{equation}
by a factor of $18$, which simply arises from the fact that we are considering scalar waves as opposed to tensor waves. 
Note however that in the models we are considering,  \eqn\eqref{eqn:kgp2_mplapp} would be the amount of power emitted in (non-Vainshtein screened) scalar waves in addition to the standard tensor gravitational waves emitted in GR. 
The total power emitted in that case would hence be enhanced by about $1/18 \approx 5\%$, as compared with GR which would be in clear disagreement with observations. 
This is yet another manifestation of the fact that a standard scalar field conformally coupled to matter with $\mpl$ strength would strongly violate the equivalence principle and is incompatible with observations. 
In what follows we shall see analytically how the Vainshtein mechanism changes this effect.

Since we are assuming that the two bodies are moving in a Keplerian orbit, Kepler's third law allows us to rewrite this as
\begin{equation}
	\label{eqn:kgp2_prapp}
	P_2^\text{KG} = \frac{M}{\bar r} \frac{( \op \bar r)^8}{45}.
\end{equation}
This particular form is useful for checking the power reported by the simulation (see \sect\ref{sec:numerics} and \app\ref{app:rescale} for more details).

\subsection{Cubic Galileon Radiation --- the effect of Vainshtein screening}
\label{sec:cubpert}
We now turn to the cubic Galileon (finite $\Lambda$) where for sufficiently small values of the strong coupling scale $\Lambda$ (\ie large Vainshtein radius $\rv$ as defined in \eqn\eqref{eqn:rv}), we expect the Vainshtein screening to be active and to suppress the power emitted in scalar waves.

To compute the power for the cubic Galileon we start by solving \eqn\eqref{eqn:bkg_form}.
This gives two branches for $E$, but as mentioned earlier we will focus on the one that provides the correct asymptotic conditions at infinity (standard Newton inverse square law). That solution is
\begin{equation}
	\label{eqn:cub_bkg}
	E = -\frac{\Lambda^3}{4}r\left[3- \sqrt{9 +\frac{32}{\pi}\left(\frac{\rv}{r}\right)^3}\right].
\end{equation}
To calculate the power, we are interested in the mode functions at $\bar r$ that is $r\ll \rv$. In this limit the components of the effective metric in \eqn\eqref{eqn:eff_comp} are,
\begin{align}
	\label{eqn:cubic_Z}
	Z^{tt}&\approx -\sqrt{\frac 2\pi}\left(\frac{\rv}r\right)^
	{3/2},&
	Z^{rr}&\approx \frac43\sqrt{\frac 2\pi}\left(\frac{\rv}r\right)^{3/2},&
	Z^{\Omega\Omega}&\approx \frac13\sqrt{\frac 2\pi}\left(\frac{\rv}r\right)^{3/2}.
\end{align}
Thus the differential equation governing the radial part of the mode functions $u_{l\omega}$ is
\begin{equation}
	\left(\omega^2-\frac{l(l+1)}{3r^2}\right)u_{l\omega}(r)+\frac{2}{3r}u_{l\omega}'(r)+\frac43u_{l\omega}''(r)=0,
\end{equation}
yielding the solutions
\begin{equation}
	u_{l\omega}=\bar u_{l\omega} r^{1/4}\left(A_{l\omega}J_{\alpha_l}(\sqrt{3}r\omega/2)+B_{l\omega} J_{-\alpha_l}(\sqrt{3}r\omega/2)\right),
\end{equation}
where the $J_\alpha$ are the Bessel functions and $\alpha_l=\frac14 (1+2l)$.
Requiring the solution to be finite at $r=0$ forces $B_{l\omega}=0$ for $l>1$. Regularizing the delta function source over a distance $\epsilon\ll \bar r$, the background field configuration alters to $E\propto \Lambda^3 \frac{r \rv}{\epsilon}$ in the $r\ll\epsilon$ limit.
Using this to determine the coefficients $A_{l,\omega}$ and $B_{l,\omega}$ we find that $A_{l,\omega}=0$ for $l=0$ and $B_{l,\omega}=0$ for $l>0$.

The normalization of the mode functions is set by \eqn\eqref{eqn:prop_def}. Using the definition of the Wightman functions \eqn\eqref{eqn:prop_wight}, the normalization condition reduces to
\begin{equation}
	\label{eqn:mode_norm}
	Z^{tt}\left.\left(\partial_t W^+(x,x')-\partial_t W^-(x,x')\right)\right|_{t=t'}=i\delta^3(\vec x-\vec x').
\end{equation}
The general form of the mode functions \eqn\eqref{eqn:permodefunc} allows us to write the time derivative of the Wightman functions as
\begin{equation}
	\left.\partial_t W^{\pm}\right|_{t=t'}=\mp i\sum_{lm}\int_0^\infty\diff{\omega}\omega u_{l\omega}(r)u_{l\omega}(r')Y_{lm}(\theta,\phi)Y_{lm}^*(\theta,\phi).
\end{equation}
Using the closure of the spherical harmonics and $\delta(r-r')/r=\int_0^\infty\diff{q} qJ_\nu(qr)J_\nu(qr')$, the Wightman functions reduce to
\begin{equation}
	\label{eqn:cubic_wightman}
	\left.\partial_t W^{\pm}\right|_{t=t'}=\mp \frac43i\bar u^2 r^{3/2}\delta^3(\vec x-\vec x').
\end{equation}
Using this to reduce the normalization condition \eqn\eqref{eqn:mode_norm} we find $\bar u^2=(3/8)\sqrt{\pi/(2\rv^3)}.$
Thus the properly normalized mode functions for the cubic Galileon are
\begin{equation}
	\phi_{lm\omega}=\left(\frac{9\pi}{128}\right)^{1/4}\left(\frac{r}{\rv^3}\right)^{1/4}J_{\alpha_l}\left(\sqrt{3}r\omega/2\right)Y_{lm}(\theta,\phi)e^{-i\omega t},
\end{equation}
where
\begin{equation}
	\alpha_l=\begin{cases}-\frac14& l=0\\\frac14(1+2l) &l>0\end{cases}.
\end{equation}
In the $r\ll \op$ limit the radial mode functions take the asymptotic form
\begin{equation}
	u_{ln}(r) \approx \left(\frac{9\pi}{128}\right)^{1/4}\left(\frac{r}{\rv^3}\right)^{1/4}\frac{(\sqrt{3}rn\op/4)^{\alpha_l}}{\Gamma(1+\alpha)}.
\end{equation}
Since $P_l\ \propto \ u_l^2 \ \propto\ (\op\bar r)^{(1+2l)/4}$ the multipole power is $l$-suppressed, meaning the quadrupole dominates the total power since the monopole and dipole power are zero (in the non-relativistic limit, see \sect\ref{sec:eff_act_pow}). The quadrupole power is given by,
\begin{equation}
	P_2^\text{cubic}=\frac{M^2}{8 \pi \mpl^2} \frac{45\times 3^{1/4} \pi^{3/2}}{1024\, \Gamma
   \left(\frac{9}{4}\right)^2} \frac{(\op\bar r)^3}{ (\op\rv)^{3/2}}\op^{2}.
\end{equation}
Comparing to the GR result we see a ratio
\begin{equation}
	\label{eqn:cu_power_ratapp}
	\frac{P^\text{cubic}_2}{P^\text{GR}_2} = \frac{225\times 3^{1/4}\pi^{3/2}}{2048\times\Gamma\left(\frac94\right)^2} (\op\bar r)^{-1} (\op\rv)^{-3/2}.
\end{equation}
Importantly the normal Vainshtein suppression of the Galileon force between the two objects goes as $(\bar r/\rv)^{3/2}$, where in this dynamic system it goes as $(\op\bar r)^{-1} (\op\rv)^{-3/2}$. This implies a weakening of the Vainshtein mechanism by the orbital velocity $v^{5/2}=(\op\bar r)^{5/2}$ in comparison to the static case.

\section{Program Rescaling}
\label{app:rescale}
In order to work with appropriate dimensionless program variables, we rescale the problem as follows:
\begin{align}
x^{\mu}&=A x_{\text{pr}}^{\mu},\\
\pi&=B\pi_{\text{pr}},\\
T&=\frac{B^2}{A^2} T_{\text{pr}}\,,
\end{align}
where $T$ is the trace of the stress energy.
We choose
\begin{align}
\label{eqn:tildr_def}
A&=\frac{\bar r}{2}=\tilde r\\
B&=\sqrt{\frac{M}{\bar r}}\,,
\end{align}
where $\bar r$ is the semi major axis and $M/2$ is the mass of each star.
We impose a Keplerian orbit on the two bodies so the orbital frequency of the system is given by
\begin{align}
\label{r:Period}
\op^{2}&=
\frac{M}{8\pi \mpl^{2}\bar r^{3}}\,.
\end{align}
The Vainshtein radius associated with this two body system is
\begin{align}
\rv=\frac1{\Lambda}\left(\frac{M}{16 \mpl}\right)^{1/3}.
\end{align}
Finally we define two dimensionless parameters $\lambda$ and $\kappa$ as
\begin{align}
\lambda&=\frac{B}{\mpl}=2 \sqrt{2\pi} \left(\op \bar r \right)\\
\kappa&=\frac{B}{A^{2}\Lambda^{3}}=\frac{32}{\sqrt{2\pi}} \frac{\left(\op \rv\right)^3}{\left(\op \bar r \right)^4}.
\end{align}
The free Klein-Gordon model (no cubic interaction, no Vainshtein) corresponds to $\Lambda \to \infty$ (\ie \, $\kappa=0$). Switching on cubic Galileon interactions (\ie \, switching on what should be the Vainshtein mechanism), corresponds to $\kappa>0$.

\end{appendix}

\bibliographystyle{JHEP}
\bibliography{refs}

\end{document}